\documentclass[12pt]{article}

\usepackage{amsmath}
\usepackage{graphicx}
\usepackage{amsfonts}
\usepackage{amssymb}
\usepackage{latexsym}
\usepackage{color}
\usepackage{amssymb}

\newcommand{\vect}[1]{\mbox{\boldmath${#1}$}}
 
\newcommand{\be}{\begin{eqnarray}} 
\newcommand{\ee}{\end{eqnarray}}

\newcommand{\beq}{\begin{equation}}
\newcommand{\eeq}{\end{equation}}
\newcommand{\beqa}{\begin{eqnarray}}
\newcommand{\eeqa}{\end{eqnarray}}

\newcommand{\lmk}{\left(}
\newcommand{\rmk}{\right)}
\newcommand{\lkk}{\left[}
\newcommand{\rkk}{\right]}

\newcommand{\la}{\langle}
\newcommand{\ra}{\rangle}

\newcommand{\vs}{{\vect s}}
\newcommand{\vx}{{\vect x}}
\newcommand{\vy}{{\vect y}}

\newcommand{\tA}{{\tilde A}}
\newcommand{\tW}{{\tilde W}}
\newcommand{\ta}{{\tilde a}}
\newcommand{\tw}{{\tilde w}}

\begin{document}

\begin{titlepage}

\begin{flushright}
original article submitted by invitation II
\end{flushright}

\begin{flushright}
RESCEU 18/16
\end{flushright}

\begin{center}


\vskip .5in

{\Large \bf
Toward the detection of gravitational waves 
under non-Gaussian noises I\hspace{-.1em}I.\\
Independent Component Analysis
}

\vskip .45in

{\large
Soichiro Morisaki$^{1,2}$,
Jun'ichi Yokoyama$^{1,2,3}$,\\
Kazunari Eda$^{1,2}$,
Yousuke Itoh$^{1}$
}

\vskip .45in

{\em
$^1$
  Research Center for the Early Universe (RESCEU), Graduate School
  of Science,\\ The University of Tokyo, Tokyo 113-0033, Japan
  }\\
{\em
$^2$
  Department of Physics, Graduate School of Science, The University of Tokyo, \\Tokyo, 113-0033, Japan  
  }\\
{\em
$^3$
  Kavli Institute for the Physics and Mathematics of the Universe (Kavli
 IPMU), UTIAS, WPI,
  The University of Tokyo, Kashiwa, Chiba, 277-8568, Japan
  }

e-mail: morisaki@resceu.s.u-tokyo.ac.jp, yokoyama@resceu.s.u-tokyo.ac.jp, eda@resceu.s.u-tokyo.ac.jp, yousuke\_itoh@resceu.s.u-tokyo.ac.jp

\end{center}

\vskip .4in

\begin{abstract}
We introduce a new analysis method to deal with stationary non-Gaussian noises in gravitational wave detectors in terms of the independent component analysis. 
First, we consider the simplest case where the detector outputs are linear combinations of the inputs, consisting of signals and various noises, and show that this method may be helpful to increase the signal-to-noise ratio.
Next, we take into account the time delay between the inputs and the outputs.
Finally, we extend our method to nonlinearly correlated noises and show that our method can identify the coupling coefficients and remove non-Gaussian noises. Although we focus on gravitational wave data analysis, our methods are applicable to the detection of any signals under non-Gaussian noises.(116 words)

\end{abstract}
Running title: Gravitational wave detection under non-Gaussian noises I\hspace{-.1em}I.
\end{titlepage}

\newpage
\section{Introduction}
Direct detection of gravitational waves (GWs) was achieved for the first time in 2015 \cite{LIGO} 
by the advanced Laser Interferometer Gravitational wave Observatory (aLIGO) \cite{TheLIGOScientific:2014jea}.
This discovery brought a great impact on science, announcing the onset of gravitational wave astronomy.
Following aLIGO, the large-scale cryogenic gravitational wave telescope (LCGT) now known as KAGRA, is being constructed in Kamioka, Japan \cite{KAGRA3}.
It is the first detector in deep underground and expected to provide useful knowledges for future detectors.
In addition, it will improve determination accuracy of a gravitational wave source direction on the sky together with aLIGO and Virgo, and help to obtain more information from GWs.
 
Because gravity is the weakest force among the four elementary 
interactions known to the present, GWs have a high penetrating power.
Thanks to the smallness of their cross section, 
GWs are hard to be influenced by interstellar medium 
during the propagation unlike electromagnetic waves. 
GWs bring information on deep inside of compact stars such as 
neutron stars directly to us. By the same token, 
this property makes its detection very difficult. 
To extract weak GW signals buried in detector noises,   
a commonly used method is the matched filtering technique 
which is based on the maximum likelihood method 
assuming stationary, Gaussian noises \cite{GAUSSIANmethod}. 
However, non-Gaussian noises frequently appear in actually measured output data.
They would hinder the sensitivity of the matched filter,
which is known to be susceptible to non-Gaussian noises 
stemming from instrumental and environmental artifacts, and results in false alarms. 
One approach to deal with non-Gaussian noises is 
to appropriately modify the functional form of the likelihood \cite{Rover:2008yp, Rover}. 
In the previous paper, one of us introduced the likelihood function based on 
the Edgeworth expansion for weak non-Gaussian noise 
and the Gaussian mapping method for strong non-Gaussian noise \cite{first}.
   
In this paper, we propose another method to mitigate the effect of 
non-Gaussian noises using the independent component analysis (ICA) \cite{ICA1,ICA2,ICA3}
(see \cite{icab1,icab2} for textbooks).  
The ICA has been developed in the context of blind source separation 
among which the cocktail party problem is well-known as a representative example. 
The ICA enables us to decompose output data into statistically independent components 
on the assumption that there is at most one Gaussian component in the data. 
Here, we apply the ICA method to data analysis for burst-like GW signals
and investigate how well the ICA works to
separate stationary non-Gaussian noise from output data. 

The rest of the paper is organized as follows. In \S 2.1, we consider the simplest case where the detector outputs are linear combinations of the inputs. Next, the time delay between the inputs and the outputs, which exists in real experiments, is taken into account in \S 2.2. Finally, we study the applicability of our method to nonlinearly correlated noises in \S 2.3. The last section \S 3 is devoted to our conclusion.     

\section{Independent Component Analysis (ICA)}

As is seen in the previous paper \cite{first}, signal detection under non-Gaussian
noises is much involved than the case with Gaussian counter parts since the optimal
statistic has a more complicated form.  Interestingly, however, there
have been some proposals to make use of non-Gaussian natures of both
signals and noises to separate
signals from non-Gaussian noises known as the 
independent component analysis (ICA).
Here we consider applicability of this new approach for the detection of
GWs in a simple model. 

Suppose that there exist $N$ statistically
independent sources of signals and $N$ outputs
and that there is at most one source which follows the Gaussian
distribution and all the other sources obey non-Gaussian
distributions.
The ICA is a method to identify 
each independent source of signals making use of the non-Gaussianity and 
statistical independence of sources.  So here non-Gaussianity is not an obstacle
of the analysis but rather a necessary ingredient.  Indeed, if 
multiple sources obey Gaussian distributions, we cannot distinguish
them using ICA even if they are statistically independent.

{\bf 2.1 The simplest model.}~~~Here let us consider a simple problem as a first step of realistic
application of ICA to the detection of GWs.
To be specific, let us identify two sources of signals $s_1(t)$,
and $s_2(t)$ as a burst-like gravitational wave signal $h(t)$ and non-Gaussian 
noise $k(t)$ such as seismic noises.  That is,
\beq
\vs(t)=\begin{pmatrix} s_1(t)\\ s_2(t)\end{pmatrix}=\begin{pmatrix} h(t)\\ k(t)\end{pmatrix}.
\eeq
In addition to the output from the laser interferometer $x_1(t)$,
we make use of the output from an environmental monitor such as a seismograph $x_2(t)$, and assume that
they are linear functions of the signal $\vs(t)$ as
\beq
\vx(t)=\begin{pmatrix} x_1(t)\\ x_2(t)\end{pmatrix}=A\vs(t)  \label{vs}
\eeq
where $A$ is assumed to be a time independent matrix.

By definition, the gravitational wave signal obeys a probability
distribution function (PDF)
\beq
 r_1(h,t)=\delta (h-h(t,\theta))
\eeq
where $h(t,\theta)$ is the actual waveform of gravitational radiation 
emitted from some source, say a binary neutron star coalescence, to be
observed at the position of a laser interferometer, where
 $\theta$ collectively denotes parameters of the source.

On the other hand, we do not specify the PDF of $k(t)$, $r_2(s_2)$, except that
it is a super-Gaussian distribution such as a Student t-distribution
with a larger tail than Gaussian \cite{Characterization}.

The detector output of a laser interferometer, of course, suffers from 
Gaussian noises $n(t)$ besides non-Gaussian noise $k(t)$.
Hence (\ref{vs}) should actually read
\beq
\vx(t)=A\vs(t)+{\vect n}(t),~~~{\vect n}(t)=\begin{pmatrix}
					    n(t)\\0\end{pmatrix}.
					    \label{linearmodel}
\eeq
Here we have not incorporated any Gaussian noise to the second line
where the signal $k(t)$ itself consists of (non-Gaussian) noises and
any Gaussian noise can be incorporated to a part of it.  

We now introduce a trick to replace the original source $s_1(t)=h(t)$
by $s_1(t)=h(t)+n(t)$, that is, we regard the Gaussian noise as a part
of the original signal.  Since $n(t)$ is a Gaussian noise with vanishing mean,
its statistical property is entirely characterized by the two-point
correlation function $K(t-t')=E[n(t)n(t')]$.  Then the
marginal distribution function of $s_1(t)$ is given by
\beq
 r_1[s_1(t)]=\frac{1}{\sqrt{2\pi}\sigma}\exp\lkk -\frac{1}{2\sigma^2}
\lmk s_1(t)-h(t,\theta)\rmk^2\rkk,~~~\sigma^2=K(0). \label{gauss}
\eeq
Thus $s_1(t)$ now satisfies a simple Gaussian distribution which is much
easier to handle with than the delta-function distribution (\ref{vs}),
and $\vs(t)$ and $\vx(t)$ are related by a simple 
formula $\vx(t)=A\vs(t)$. Now our tentative goal is to find the inverse
matrix of $A$ whose components are not known precisely.
One may set it as
\beq
A=\begin{pmatrix} a_{11}& a_{12}\\ 0 & a_{22}\end{pmatrix}, \label{aform}
\eeq
since the gravitational wave is so weak that it will not affect any
seismograph.  

The aim of ICA is to find a linear transformation
\beq
  \vy = W\vx,
\eeq
such that each component of the transformed variables $\vy$
is mutually statistically independent. 
Thanks to the assumption (\ref{aform}), the matrix $W$ also takes a form
\beq
  W=\begin{pmatrix} w_{11}& w_{12}\\ 0 & w_{22}\end{pmatrix}. \label{wform}
\eeq
If we knew all the components of $A$, the matrix $W$ could simply be given
by the inverse matrix $W=A^{-1}$, in which case we would find
$\vy =\vs$.  However, since we do not know them
we attempt to determine $W$ in such a way that the components of $\vy$,
$y_1(t)$ and $y_2(t)$ to be statistically independent with each other as much as
possible.

The mutual independence of statistical variables may be judged by 
introducing a cost function $L(W)$ which represents a ``distance'' in the 
space of statistical distribution functionals.  As an example, we 
adopt the Kullback-Leibler divergence \cite{KL} defined between two arbitrary
PDFs $p(\vy)$ and $q(\vy)$ as 
\beq
 D[p(\vy); q(\vy)]=\int p(\vy)\ln \frac{p(\vy)}{q(\vy)}dy
=E_p\lkk \ln \frac{p(\vy)}{q(\vy)}\rkk, 
\eeq
where $E_p[\cdot]$ denotes an expectation value with respect to a PDF $p$.

We examine the distance between the real distribution function of
statistically independent variables $\vs$,
$r(\vs)=r_1[s_1(t)]r_2[s_2(t)]$,
and a distribution of $\vy$, $p_y$, 
constructed from the observed distribution function of $\vx$ through
the linear transformation $\vy=W\vx$ as 
\beq
 p_y(\vy)\equiv ||W^{-1}||p_x(\vx),
\eeq
where $||W^{-1}||$ denotes the determinant of $W^{-1}$.

The cost function of $p_y(\vy)$ from $r(\vs)$ is given by
\begin{align}
  L_r(W) &= D[p_y(\vy);r(\vy)]=E_{p_y}[\ln p_y(\vy)]-E_{p_y}[\ln r(\vy)]
   \nonumber \\
&= \int ||W^{-1}||p_x(\vx)\ln \lkk ||W^{-1}||p_x(\vx)\rkk dy 
-E_{p_y}[\ln r(\vy)] \nonumber \\
&= -\ln ||W|| + \int   p_x(\vx)\ln \lkk p_x(\vx)\rkk dx
- E_{p_y}[\ln r(\vy)] \nonumber \\
&= -H[x]- E_{p_y}[||W||\ln r(\vy)] = -H[x]- E_{p_x}[\ln p(\vx,W)],  \label{56}
\end{align}
with
\begin{equation}
p(\vx,W) \equiv ||W|| r(\vy), 
\end{equation}
and
\beq
H[x]\equiv -\int   p_x(\vx)\ln \lkk p_x(\vx)\rkk dx.
\eeq

The PDF of $\vx$ in the last expression of (\ref{56})
has $W$ dependence because $p(\vx,W)$ is
a PDF of $\vx$ which is made out of the PDF of $\vy$ ($=\vs$ in this
particular case) through the relation $\vy=W\vx$.
The above formula shows that the matrix $W$ which minimizes the cost
function $L_r(W)$ also maximizes the log-likelihood ratio of $\vx$. 

Since we do not know $r(\vy)$ a priori, we instead adopt an arbitrary 
mutually independent distribution $q(\vy)=q_1(y_1)q_2(y_2)$
in the cost function.
Defining a PDF consisting of marginal distribution functions
\beq
 \tilde{p}(\vy)\equiv \int p_y(y_1,y_2)dy_2\int p_y(y_1,y_2)dy_1=
\tilde{p}_{1}(y_1)\tilde{p}_{2}(y_2),
\eeq
we find the following relation
\beq
L_q(W)=D[p_y(\vy);q(\vy)]=D[p_y(\vy);\tilde{p}(\vy)]+
D[\tilde{p}(\vy);q(\vy)]
\eeq
holds.  Since the Kullback-Leibler divergence is known to be
positive semi-definite, a distribution that minimizes the first term
in the right-hand-side yields the desired linear transformation
$\vy=W\vx$ for which this term vanishes.  In this case the second
term gives a discrepancy due to the possible incorrect choice of $q(\vy)$.
In this sense it would be better to choose a realistic trial function
$q(\vy)$ as much as possible.

It is known in fact that even for an arbitrary choice of $q(\vy)$,
the correct $W$ gives an extremum of $L_q(W)$.
Hence we solve
\beq
  \frac{\partial L_q(W)}{\partial w_{ij}}=0. \label{partial}
\eeq
  From 
\beq
L_q(W)=-H[x]-\ln||W||-E_{p_y}[\ln q(\vy)]\equiv
-H[x]-\ln||W||-E_{p_y}[f(\vy)],
\eeq
\beq
f(\vy) \equiv \ln q(\vy), 
\eeq
\begin{align}
 d_W\ln||W|| &\equiv \ln||W+dW||-\ln||W||=\ln||{\vect 1}+dWW^{-1}|| \nonumber \\
&={\rm Tr}(dWW^{-1})=(W^{-1})_{ji}dw_{ij}, \nonumber
\end{align}
and
\begin{align}
 d_W f(\vy) &\equiv f\lmk (W+dW)\vx\rmk -f(W\vx)
=\frac{\partial f}{\partial y_i}dw_{ij}x_j, \nonumber
\end{align}
we find
\begin{align}
d_W L_q(W)=E_{p_y}\lkk -(W^{-1})_{ji}-x_j\frac{\partial f}{\partial
 y_i}\rkk dw_{ij}.
\end{align}
In order to satisfy (\ref{partial}) the above expectation value should
vanish for each index.
Multiplying $w_{kj}$ to the argument of the expectation value, we find
it equivalent to
\beq
  E_{p_y}\lkk y_k\frac{\partial f}{\partial y_i}\rkk +\delta_{ki}=0 .
\eeq
That is, we require
\beq
 E_{p_y}[\varphi_i(y_i)y_j]=\delta_{ij}
\eeq
with 
\beq
 \varphi_i(y_i)\equiv -\frac{d~}{dy_i}\ln q_i(y_i). \label{independence}
\eeq

We determine $W$ so that (\ref{independence}) is satisfied for each
component choosing plausible forms of $q_i(y_i)$.  For $q_1(y_1)$ we take
\beq
  q_1(y_1)=\frac{1}{\sqrt{2\pi}\sigma}\exp\lkk -
  \frac{(y_1-h(t,\theta))^2}{2\sigma^2}\rkk ,
\eeq
based on  (\ref{gauss}), so that
$\varphi_1(y_1)=(y_1-h(t,\theta))/\sigma^2$.
In real experiments, we do not know $h(t,\theta)$ a priori. However, as found later, when we take temporal average, the contributions from gravitational waves can be neglected. Therefore, we can set $h(t,\theta)=0$ when we apply $q_1(y_1)$ to real analysis.  
As for $\varphi_2(y_2)$, it is recommended to take
\beq
 \varphi_2(y_2)=c_2\tanh y_2
\eeq
to model a super-Gaussian distribution \cite{SA}.
Using these expressions in (\ref{independence}) we determine $W$ 
which relates each component of $\vy$ and $\vx$ as
$y_1=w_{11}x_1+w_{12}x_2$ and $y_2=w_{22}x_2$.
In doing so we replace the ensemble average $E[\cdot]$ by temporal 
average of observed values of $\vx$ which we denote by brackets.

Each component of (\ref{independence}) reads as follows.
\begin{align}
1&=
 \sigma^{-2}E[(w_{11}x_1+w_{12}x_2-h)(w_{11}x_1+w_{12}x_2)]\nonumber \\
&=\sigma^{-2}\lkk w_{11}^2\la x_1^2\ra +2w_{11}w_{12}\la x_1x_2\ra
+w_{12}^2\la x_2^2\ra -(w_{11}\la h x_1\ra+ w_{12}\la h x_2\ra)\rkk, \label{11}\\
0&=\sigma^{-2}E[(w_{11}x_1+w_{12}x_2-h)w_{22}x_2] \nonumber \\
&= \sigma^{-2}\lkk w_{11}w_{22}\la x_1x_2\ra +w_{12}w_{22}\la x_2^2\ra
-w_{22}\la h x_2\ra\rkk, \label{12}\\
0&=c_2 E[y_1 \tanh y_2]=c_2 E\lkk (w_{11}x_1+w_{12}x_2)\tanh
 w_{22}x_2\rkk\nonumber \\
&= c_2w_{11} \la x_1\tanh (w_{22}x_2) \ra +c_2 w_{12}\la x_2
\tanh( w_{22}x_2)\ra, \label{21}\\
1&=c_2 E[y_2\tanh y_2]=c_2 w_{22}\la x_2 \tanh( w_{22}x_2) \ra. \label{22}
\end{align}

Because a gravitational wave with a detectable amplitude is a rare
event, long-time averages of $h x_1$ and $h x_2$ should vanish. 
Then from (\ref{12}) we obtain
\beq
  w_{12}= -\frac{\la x_1x_2\ra}{\la x_2^2 \ra}w_{11},  \label{1211}
\eeq
and from (\ref{11})
\beq
 \sigma^2=w_{11}^2\la x_1^2\ra +w_{11}w_{12}\la x_1 x_2\ra,
\eeq
so that 
\beq 
 w_{11}=\lmk\frac{\la x_2^2\ra}{\la x^2_1\ra \la x_2^2\ra - \la x_1
 x_2\ra^2}\rmk^{\frac{1}{2}}\sigma~~~{\rm and}~~
w_{12}=-\frac{\la x_1x_2\ra \sigma}{(\la x_1^2\ra \la x_2^2\ra^2-
\la x_1x_2\ra \la x_2^2\ra)^{\frac{1}{2}}}. \label{1112}
\eeq

On the other hand, (\ref{21}) yields a relation
\beq
  w_{12}=-\frac{\la x_1\tanh(w_{22}x_2)\ra}{\la
  x_2\tanh(w_{22}x_2)\ra}w_{11}.\label{tanh}
\eeq
It will be found later that this condition is consistent with (\ref{1211}).

We now apply the likelihood ratio test to the above result.
The output of a laser interferometer $x_1$ consists of
\beq
 x_1=a_{11}s_1+a_{12}s_2=a_{11}h+a_{11}n+a_{12}k.  \label{78}
\eeq
If there were only Gaussian noises with $a_{12}=0$,
we would find 
\[
 E_{p_x}[x_1]  =a_{11}h,~~ E_{p_x}[x_1^2]  - (E_{p_x}[ x_1 ])^2=
a_{11}^2E[n^2]
\]
so that
\beq
 \frac{S}{N}=\frac{a_{11}h}{a_{11}\sqrt{E[ n^2]}}
=\frac{h}{\sqrt{E[n^2]}}.
\eeq
In the presence of non-Gaussian noise
instead we find 
\beq
 E_{p_x}[ x_1^2 ] - (E_{p_x}[ x_1 ])^2=a_{11}^2E[ n^2 ]+a_{12}^2 E[ k^2]
\eeq
assuming $E[ nk ] =0$.  As a result signal to noise ratio ($S/N$) gets worse
\beq
 \frac{S}{N}=\frac{h}{\sqrt{E[ n^2]+\frac{a_{12}^2}{a_{11}^2}
 E[k^2]}}.
\eeq

Now we calculate $S/N$ of $y_1$ variable from
\beq
E_{p_y}[ y_1 ] =w_{11}E_{p_x}[ x_1] +w_{12}E_{p_x}[ x_2] =w_{11}a_{11}h
,\nonumber
\eeq\beq
E_{p_y}[ y_1^2 ] -(E_{p_y}[ y_1])^2=w_{11}^2a_{11}^2E[ n^2] +
(w_{11}a_{12}+w_{12}a_{22})^2E[ k^2] .\nonumber
\eeq
If $W$ is solved exactly, it should be identical to the inverse matrix
of $A$, namely,
\beq
 \begin{pmatrix} w_{11}&w_{12}\\ 0&w_{22}\end{pmatrix}
=\begin{pmatrix} \frac{1}{a_{11}}& -\frac{a_{12}}{a_{11}a_{22}}\\
0& \frac{1}{a_{22}}\end{pmatrix},
\eeq
then we find
\beq
E_{p_y}[ y_1 ] =w_{11}E_{p_x}[ x_1] +w_{12}E_{p_x}[ x_2] =h, \nonumber
\eeq\beq
E_{p_y}[ y_1^2] -(E_{p_y}[ y_1])^2=E[ n^2].  \nonumber
\eeq

But in fact we cannot hope to obtain $W$ as an inverse matrix, but
instead
estimate it from temporal average of the observed sample.
From (\ref{1211}), (\ref{tanh}), and (\ref{78}), we find
\beq
w_{12}=-\frac{\la x_1x_2\ra}{\la
  x_2^2\ra}w_{11} \approx -\frac{a_{12}a_{22}\la
  k^2\ra}{a_{22}^2\la k^2\ra}w_{11}=-\frac{a_{12}}{a_{22}}w_{11},
\eeq
and
\begin{eqnarray}
w_{12} &=& -\frac{\la x_1\tanh(w_{22}x_2)\ra}{\la
  x_2\tanh(w_{22}x_2)\ra}w_{11} \nonumber \\
  &\approx&
  -\frac{ a_{12} \la k \tanh(w_{22}a_{22}k) \ra }{a_{22} \la k \tanh(w_{22} a_{22} k) \ra} w_{11}\nonumber \\
  &=&
  -\frac{a_{12}}{a_{22}} w_{11},
 \end{eqnarray} 
 which means these two equations are consistent with each other.
We also find that, even from the observational data, we can deduce
$w_{11}a_{12}+w_{12}a_{22}=0$ and $y_1$ variable is indeed free from
non-Gaussian noise.  The resultant $S/N$ of $y_1$ is
simply given by
\beq
  \frac{S}{N} \approx \frac{h}{\sqrt{E[ n^2]}}.
\eeq
Thus, the non-Gaussian noise is effectively removed here.

{\bf 2.2 memory effect.}~~~The above simple model (\ref{vs}) is just the first step to analyze
realistic detectors.  In fact, $x_1(t)$ would depend not only on 
$s_2(t)$ but also on some retarded times as well thanks to  operation
of a sophisticated anti-vibration system.
In order to incorporate such a memory effect, we employ the
following
model as the second step.
\beq
   \vx(t)=\sum_{\tau=0}^{\Theta-1} A(\tau)\vs(t-\tau). \label{memory}
\eeq
In this subsection, all the time variables are dimensionless
being expressed in unit of the measurement interval.
The above expression corresponds to
a model such that $\vx(t)$ depends on $\vs(t)$ up to $\Theta$ time steps before
the measurement time $t$. 
That is, we assume that the correlation between $\vx(t)$ and $\vs(t-\tau)(\tau\ge\Theta)$ is negligible.

Taking data from $t=t_s$ to $t=t_s+T-1$, $\left\{\vx(t)\lvert t=t_s,t_s+1,...,t_s+T-1\right\}$, and working in the Fourier space 
\beq
 \vx(t)=\frac{1}{T} \sum_{N=0}^{T-1}\tilde{\vx}(\omega_N;t_s)e^{i\omega_N t},~~~
 \omega_N=\frac{2\pi}{T}N,
\eeq
we find
\begin{align}
 \tilde{\vx}(\omega_N;t_s)&=\sum_{t=t_s}^{t_s+T-1}\vx(t)
e^{-i\omega_N t}
\nonumber \\
&=\sum_{t=t_s}^{t_s+T-1}\sum_{\tau=0}^{\Theta-1}A(\tau)\vs(t-\tau)e^{-i\omega_N t} \nonumber \\
&=\sum_{\tau=0}^{\Theta-1}A(\tau)e^{-i\omega_N \tau} \sum_{t=t_s}^{t_s+T-1}\vs(t-\tau)e^{-i\omega_N(t-\tau)}. 
\end{align}
If $T \gg \Theta$, the sum over $t$ can be approximated by
\beq
\sum_{t=t_s}^{t_s+T-1}\vs(t-\tau)e^{-i\omega_N(t-\tau)} \thickapprox \sum_{t=t_s}^{t_s+T-1}\vs(t)e^{-i\omega_N t} = \tilde{\vs}(\omega_N;t_s).
\eeq
Thus, when we take a long time-series compared to $\Theta$ for the Fourier expansion, the following equality holds for each Fourier mode.
\beq
\tilde{\vx}(\omega_N;t_s)=\tilde{A}(\omega_N)\tilde{\vs}(\omega_N;t_s),~~~
\tilde{A}(\omega_N) \equiv \sum_{\tau=0}^{\Theta-1}A(\tau)e^{-i\omega_N \tau}.
\eeq
To derive this relation, we have assumed that $A$ is independent of $t_s$.

Then for each Fourier mode, $A$ and $W$ take the following form
\beq
\tA(\omega_N)=\begin{pmatrix} \ta_{11}(\omega_N) & \ta_{12}(\omega_N) \\
0 & \ta_{22}(\omega_N)\end{pmatrix},~~~
\tW(\omega_N)=\begin{pmatrix} \tw_{11}(\omega_N) & \tw_{12}(\omega_N) \\
0 & \tw_{22}(\omega_N)\end{pmatrix}.
\eeq
Since the normalization of $y_i$ is arbitrary, one can put
$\tw_{11}(\omega_N)=\tw_{22}(\omega_N)=1$.
Calculating the Fourier components for various values of $t_s$, for example $t_s=0,T,2T,...,MT (\mathrm{M~is~an~integer.})$, and
following the same argument
as in the case of the simplest model, we find
\beq
 \tw_{12}(\omega_N)=-\frac{\la x_1(\omega_N;t_s)x_2(\omega_N;t_s)\ra_{t_s}}
{\la x_2^2(\omega_N;t_s)\ra_{t_s}},~~~
\tw_{11}(\omega_N)=-\frac{\la x_1(\omega_N;t_s)x_2(\omega_N;t_s)\ra_{t_s}}
{\la x_2^2(\omega_N;t_s)\ra_{t_s}} \label{w11}
\eeq
where $\la \cdot \ra_{t_s}$ denotes an average with respect to $t_s$.

So the variable
\beq
 \tilde{y}_1(\omega_N;t_s)=\tilde{x}_1(\omega_N;t_s)+\tw_{12}(\omega_N)\tilde{x}_2(\omega_N;t_s)
\eeq
follows a Gaussian distribution, and we can find the increased $S/N$
for each frequency $f=\omega/2\pi$ as in the previous subsection.

{\bf 2.3 Nonlinear coupling.}~~~So far we have considered the linear models (\ref{linearmodel}) and (\ref{memory}). However, in real gravitational wave experiments, there exist non-linearly correlated noises \cite{nonlinearnoise}. In order to investigate the applicability of ICA to nonlinear cases, we consider the following simple nonlinear model as a first step:
\beq
\begin{pmatrix} x_1(t)\\ x_2(t)\end{pmatrix}=\begin{pmatrix}a& b\\ 0 & 1\end{pmatrix}\begin{pmatrix}h(t)+n(t)\\ k(t)\end{pmatrix} + \begin{pmatrix} c[h(t)+n(t)]k(t) \\ 0 \end{pmatrix} \label{NLcoupling}
\eeq
Without loss of generality, we can set the covariances of $n(t)$ and $k(t)$ to be unity by redefining $a$, $h(t)$, $b$, and $x_2(t)$:
\beq
E[n^2(t)]=E[k^2(t)]=1 \label{covariance} ,
\eeq
and we define $n(t)$ so that
\beq
a \ge 0. \label{a_condition}
\eeq
In addition, we consider the case where $|b|$ is not much larger than $a$ and $|c|$ (see the argument above (\ref{alpha_evo2}) and (\ref{gamma_evo2})).
By assumption, the marginal PDF of $n(t)$ is the normal distribution,
\beq
 q_1(n)=\frac{1}{\sqrt{2\pi}}\exp\left( -
  \frac{n^2}{2}\right),
\eeq
whereas the PDF of $k$, $q_2(k)$, is not known.

Our goal is to remove the non-Gaussian noise $k(t)$.
If we know the coefficients, $a$, $b$, and $c$, we can obtain the time series without non-Gaussian noise, $h(t)+n(t)$ by using the transformation
\beq
h(t)+n(t)=\frac{x_1(t)-b x_2(t)}{a + c x_2(t)}. \label{removeNG}
\eeq
Therefore we estimate the values of $a$, $b$ and $c$.

When we estimate them, we consider a long-time average of the data. Assuming that the detectable burst gravitational waves rarely come to our detectors, we can neglect them. Therefore we may consider a simplified model
\beq
\begin{pmatrix} x_1(t)\\ x_2(t)\end{pmatrix}=\begin{pmatrix}a& b\\ 0 & 1\end{pmatrix}\begin{pmatrix}n(t)\\ k(t)\end{pmatrix} + \begin{pmatrix} cn(t)k(t) \\ 0 \end{pmatrix} \label{NLcoupling2}
\eeq
to estimate the coefficients.
To begin with, the coefficient $b$ can be estimated easily by
\beq
b_{est} = \la x_1(t)x_2(t) \ra, \label{b_est}
\eeq
because 
\begin{equation}
 E[n(t)k(t)]=E[n(t)]E[k(t)]=0 ,
\end{equation}
\begin{equation}
E[n^2(t)k(t)] = E[n^2(t)]E[k(t)] =0 .
\end{equation}
The next targets are $a$ and $c$.  
To find their values, we consider a transformation similar to (\ref{removeNG}), 
\beq
y_1(t)=\frac{x_1(t)-b_{est} x_2(t)}{\alpha + \gamma x_2(t)},~~~y_2(t)=x_2(t) \label{NLtransf}
\eeq
and regard $y_1(t)$ as the reconstructed $h(t)+n(t)$. They are very close to each other when $\alpha = a$ and $\gamma = c$. As the Jacobian of the transformation (\ref{NLtransf}) is given by
\beq
 J = \left| \frac{\partial (y_1,y_2)}{\partial (x_1,x_2)} \right| = \left| \frac{1}{\alpha + \gamma x_2} \right|,
 \eeq 
the cost function can be expressed as
\begin{eqnarray}
L(\alpha, \gamma) &=& D[p_y(\vy);q_1(y_1)q_2(y_2)] \nonumber \\
&=&E_{p_y}\left[\ln\left(\frac{p_y(\vy)}{q_1(y_1)q_2(y_2)}\right)\right] \nonumber \\
&=&E_{p_x}\left[\ln\left(\frac{J^{-1}p_x(\vx)}{q_1(\frac{x_1-b_{est} x_2}{\alpha + \gamma x_2})q_2(x_2)}\right)\right] \nonumber \\
&=&\frac{1}{2}E_{p_x}\left[\left(\frac{x_1-b_{est} x_2}{\alpha + \gamma x_2}\right)^2\right]+E_{p_x}\left[\ln|\alpha + \gamma x_2|\right] + \mathrm{const.}. \label{NLcost}
\end{eqnarray}
We minimize it with respect to $\alpha$ and $\gamma$. The derivatives of the cost function read
\beq
\frac{\partial L}{\partial \alpha} = E_{p_x}\left[ \frac{(\alpha + \gamma x_2)^2 - (x_1 - b_{est} x_2)^2}{(\alpha + \gamma x_2)^3} \right], \label{alpha1}
\eeq
\beq
\frac{\partial L}{\partial \gamma} = E_{p_x}\left[ \frac{x_2(\alpha + \gamma x_2)^2 - x_2(x_
1 -  b_{est} x_2)^2}{(\alpha  + \gamma x_2)^3} \right], \label{gamma1}
\eeq
and their learning rules are $\Delta \alpha \propto - \frac{\partial L}{\partial \alpha}$ and  $\Delta \gamma \propto - \frac{\partial L}{\partial \gamma}$. However, the cost function is mathematically ill-defined because of the singularity at $\alpha + \gamma x_2 = 0$. Therefore we must improve these rules. We consider the following modified learning rules,
\beq
\Delta \alpha \propto - E_{p_x}[(\alpha + \gamma x_2)^2 - (x_1 - b_{est} x_2)^2], \label{alpha2}
\eeq
\beq
\Delta \gamma \propto -E_{p_x}[x_2(\alpha + \gamma x_2)^2 - x_2(x_1 - b_{est} x_2)^2],
\label{gamma2}
\eeq
which are obtained by removing the denominators of (\ref{alpha1}) and (\ref{gamma1}). From (\ref{NLcoupling}) and (\ref{covariance}), the rules can be written as
\beq
\Delta \alpha \propto -[\alpha^2 - a^2 - (b-b_{est})^2 + \gamma^2 - c^2],  \label{alpha_evo}
\eeq
\beq
\Delta \gamma \propto -2(\alpha \gamma - ac) + [(b - b_{est})^2 - \gamma^2 +c^2] \epsilon, \label{gamma_evo}
\eeq 
where we have defined $\epsilon$ by $\epsilon = E_{p_x}[k^3(t)]$.
Now we assume $|b - b_{est}| \ll a,|c|$, which means that we do not consider the case where $|b|$ is too large and the number of the samples is too small for us to extract the information about $a$ and $c$. In this case, (\ref{alpha_evo}) and (\ref{gamma_evo}) can be written as
\beq
\Delta \alpha \propto - [ \alpha^2 + \gamma^2 - a^2 - c^2], \label{alpha_evo2}
\eeq
\beq
\Delta \gamma \propto - [ 2(\alpha \gamma - ac) + \epsilon (\gamma^2 - c^2)], \label{gamma_evo2}
\eeq	
respectively.
We study these learning rules and show that they tell us where the point ($a$,$c$) is.

\begin{figure}
\begin{center}
\includegraphics[width = 10cm]{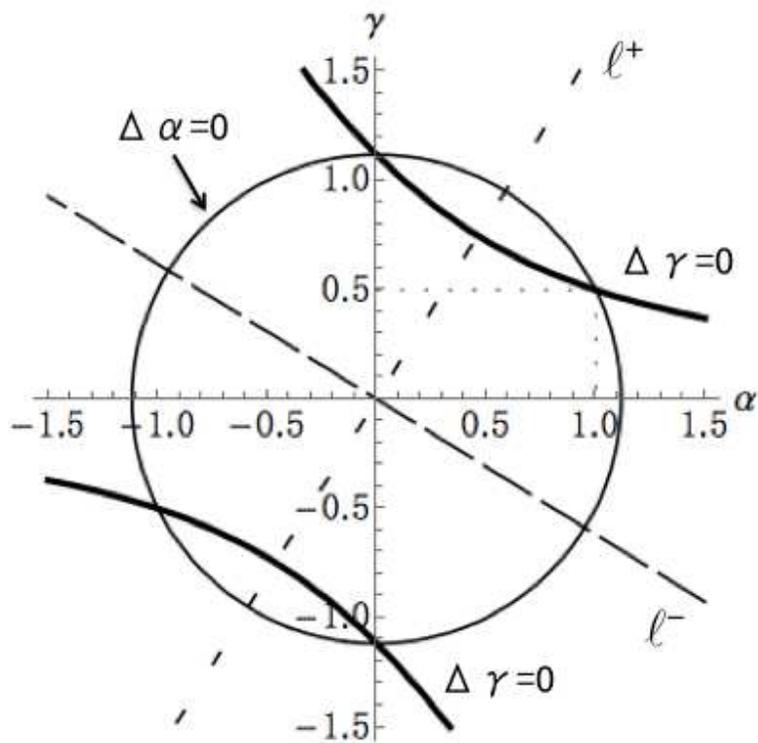}
\caption{The curves defined by $\Delta \alpha = 0$ and $\Delta \gamma = 0$ with $a=1$, $c=0.5$, $\epsilon =1$. They are symmetrical with respect to $\ell^+$ and $\ell^-$. The intersections are the stationary points of (\ref{alpha_evo2}) and (\ref{gamma_evo2}). The convergence point is one of them.}
\label{POI}
\end{center}
\end{figure}

First, we locate their stationary points, which satisfy $\Delta \alpha = 0$ and $\Delta \gamma = 0$. The curves defined by $\Delta \alpha = 0$ and $\Delta \gamma = 0$ are depicted in Fig. \ref{POI}. They are symmetrical with respect to the lines $\ell^{\pm}$ defined by the following equations,
\beq
\ell^+: \gamma = \left(\frac{(\epsilon^2 + 4)^{\frac{1}{2}} + \epsilon}{(\epsilon^2 + 4)^{\frac{1}{2}} - \epsilon} \right)^{\frac{1}{2}} \alpha, \label{l+}
\eeq
\beq
\ell^-: \gamma = - \left(\frac{(\epsilon^2 + 4)^{\frac{1}{2}} - \epsilon}{(\epsilon^2 + 4)^{\frac{1}{2}} + \epsilon} \right)^{\frac{1}{2}} \alpha. \label{l-}
\eeq
We can easily check that $(\alpha, \gamma) = (a,c)$ satisfies $\Delta \alpha = \Delta \gamma = 0$. Therefore, the points satisfying $\Delta \alpha = \Delta \gamma = 0$ are $(a,c)$ and those which are symmetrical to $(a,c)$ with respect to $\ell^+$, $\ell^-$ and the origin. 

\begin{figure}
  \begin{center}
    \begin{tabular}{c}

      \begin{minipage}{0.5\hsize}
        \begin{center}
          \includegraphics[clip, width=7cm]{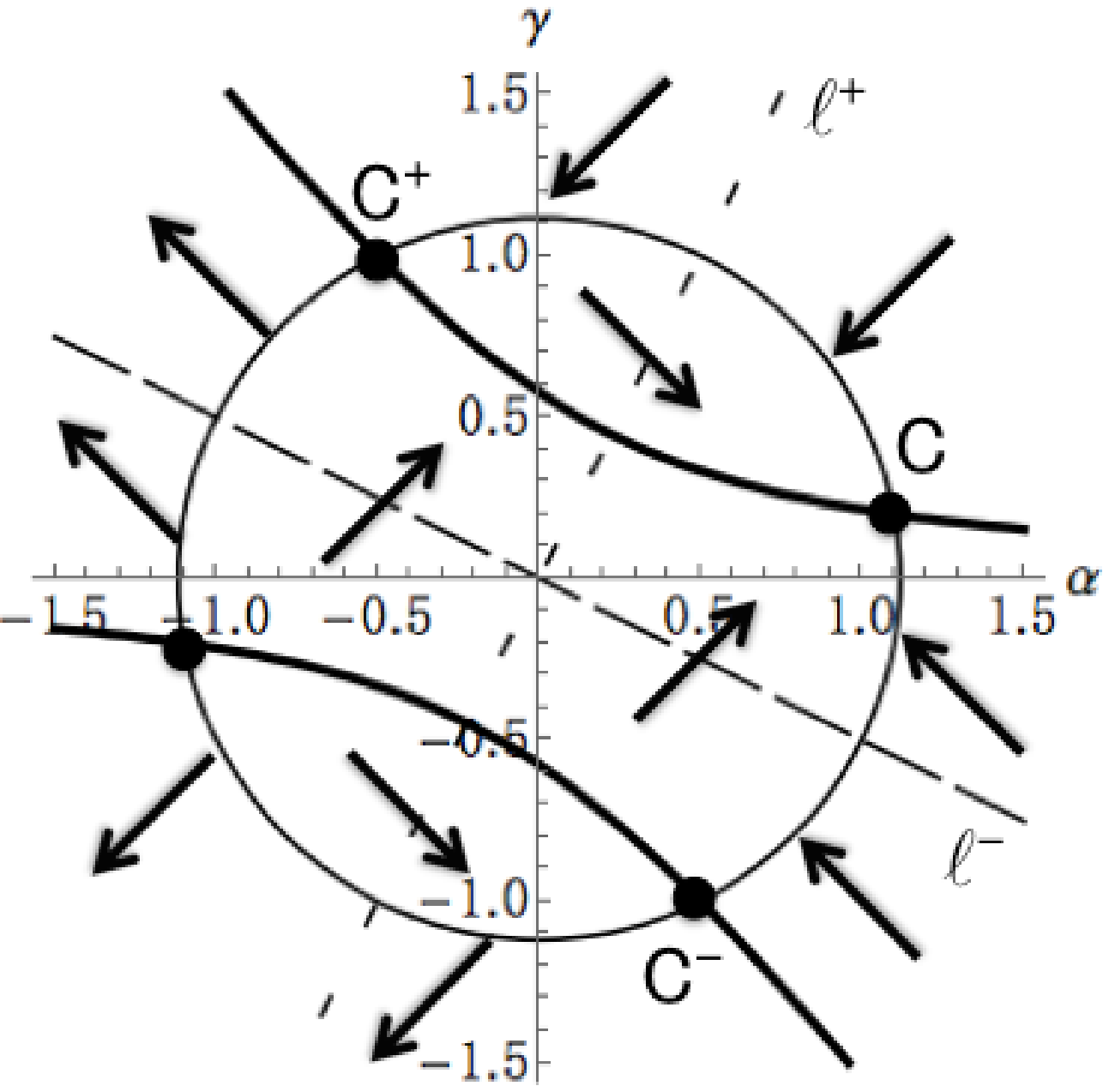}
          \hspace{1.6cm} [a]$2ac + \epsilon c^2 > 0$ \\($a = 0.5$, $c = -1$, $\epsilon = 1.5$) 
        \end{center}
      \end{minipage}

      \begin{minipage}{0.5\hsize}
        \begin{center}
          \includegraphics[clip, width=7cm]{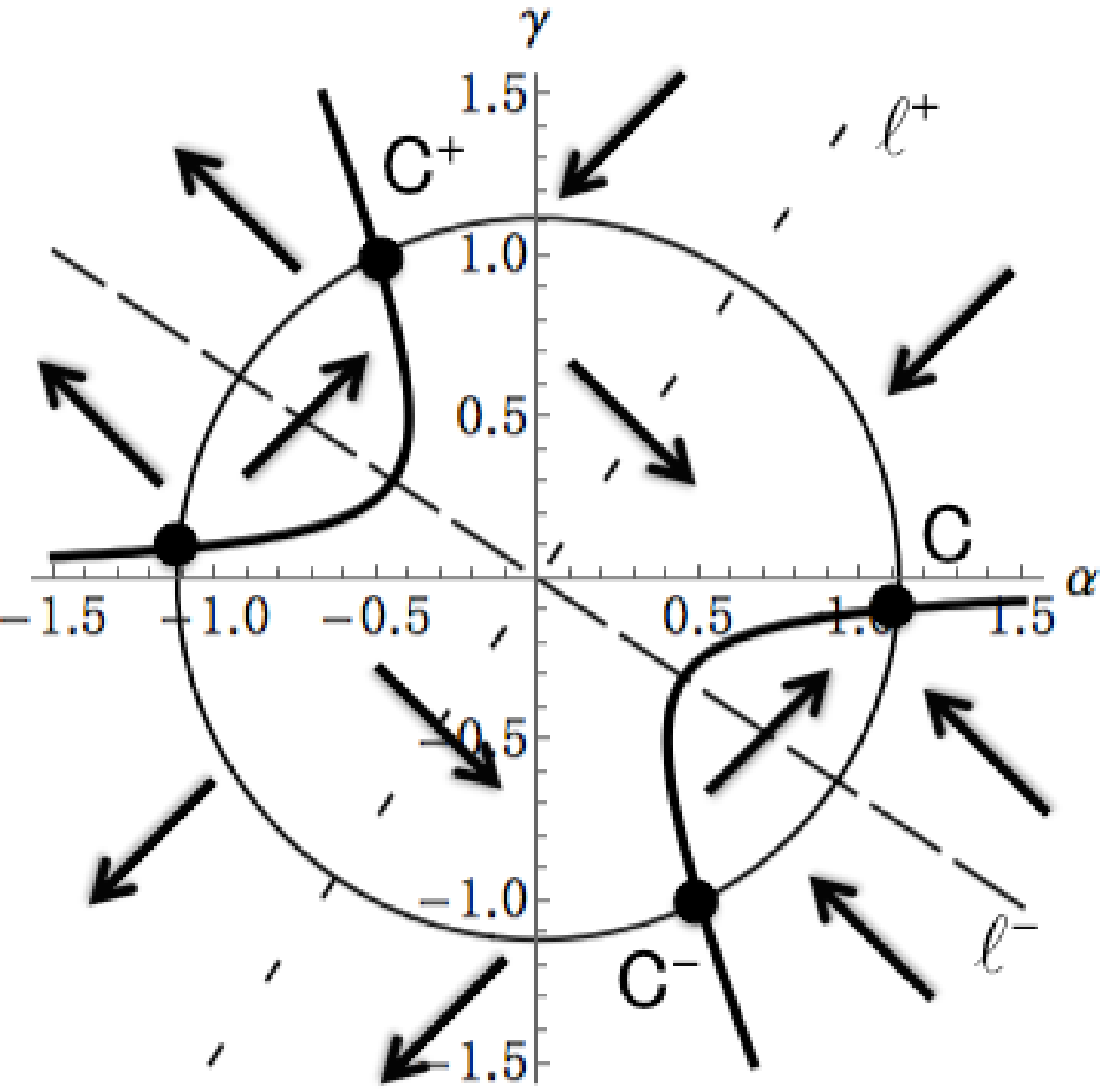}
          \hspace{1.6cm} [b]$2ac + \epsilon c^2 < 0$ \\($a = 0.5$, $c = -1$, $\epsilon = 0.8$) 
        \end{center}
      \end{minipage}

    \end{tabular}
    \caption{The flow of (\ref{alpha_evo2}) and (\ref{gamma_evo2}). It is drawn only by taking the signs of (\ref{alpha_evo2}) and (\ref{gamma_evo2}) into account, which means the lengths of the arrows are not important. The large dots represent the stationary points and the points other than C are not stable points. It can be seen that points in the region satisfying (\ref{REGION}) approach C.}
    \label{FLOW}
  \end{center}
\end{figure}

However, all these points are not necessarily stable points. In order to locate the convergence point of (\ref{alpha_evo2}) and (\ref{gamma_evo2}), we must also study the evolution of $(\alpha, \gamma)$. Focusing on the sign of $\Delta \alpha$ and $\Delta \gamma$, we depict the flow schematically in Fig. \ref{FLOW}. We refer to the stationary point on the region defined by
\beq
- \left(\frac{(\epsilon^2 + 4)^{\frac{1}{2}} - \epsilon}{(\epsilon^2 + 4)^{\frac{1}{2}} + \epsilon} \right)^{\frac{1}{2}} \alpha < \gamma <  \left(\frac{(\epsilon^2 + 4)^{\frac{1}{2}} + \epsilon}{(\epsilon^2 + 4)^{\frac{1}{2}} - \epsilon} \right)^{\frac{1}{2}} \alpha \label{REGION}
\eeq
as C$(\alpha_c,\gamma_c)$, and to the symmetrical points to C with respect to $\ell^{\pm}$ as $\mathrm{C}^{\pm}$. We can easily find that the stationary points other than C are not stable. For example, if the point $(\alpha, \gamma)$ goes in the negative direction of $\alpha$-axis from $\mathrm{C}^{\pm}$, it goes to infinity. Therefore, the point $(\alpha, \gamma)$ never converges to them. On the other hand, C is a stable point. In order to show that, we substitute $\alpha = \alpha_c + \delta \alpha$ and $\gamma = \gamma_c + \delta \gamma$ into (\ref{alpha_evo2}) and (\ref{gamma_evo2}), and keep leading terms, to find
\beq
\begin{pmatrix} \Delta (\delta \alpha)\\ \Delta (\delta \gamma)\end{pmatrix} = -2 A \begin{pmatrix} \delta \alpha\\ \delta \gamma\end{pmatrix},
\eeq
\beq
\mathrm{with} ~~~~ A = \begin{pmatrix} \alpha_c &   \gamma_c \\ \gamma_c &   \alpha_c + \epsilon \gamma_c \end{pmatrix}.
\eeq 
From (\ref{REGION}), we can easily show that the matrix $A$ has only positive eigenvalues and so C is a stable point. 
In addition, if the point $(\alpha, \gamma)$ is constrained on the region defined by (\ref{REGION}), it does not go to infinity and tends to approach C. 
Because $\epsilon$ can be estimated from the data, we can return the point onto that region by displacing it with respect to $\ell^{\pm}$ or the origin.
Therefore, with the point $(\alpha,\gamma)$ displaced properly while it is evolving, it converges to C.

\begin{figure}
	\begin{center}
		\includegraphics[width = 8cm]{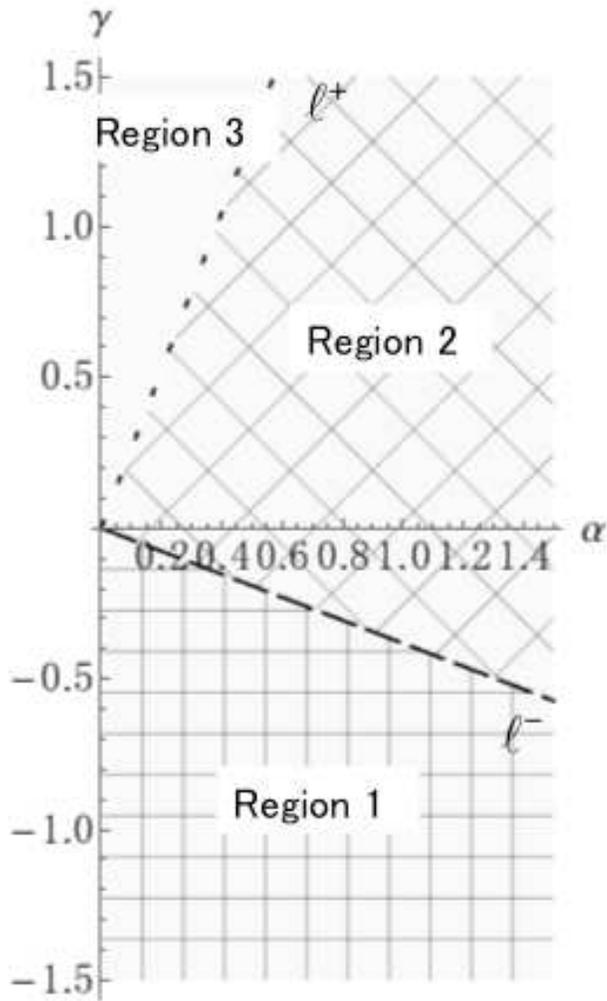}
		\caption{The right-half plane devided by $\ell^+$ and $\ell^-$. From the condition (\ref{a_condition}), $(a,c)$ is on one of these regions. On the other hand, the convergence point C is on the Region 2. Therefore whether the point $(\alpha,\gamma)$ converges to $(a,c)$ depends on which region $(a,c)$ is on.}
	\label{where}
	\end{center}
\end{figure}

Because C is just the stationary point on the region defined by (\ref{REGION}) and not necessarily coincide with $(a,c)$, there are three patterns depending on the location of $(a,c)$ shown in Fig. \ref{where}.
\begin{itemize}
\item Region 1: converges to the point symmetrical to $(a,c)$ with respect to $\ell^-$.
\item Region 2: converges to $(a,c)$.
\item Region 3: converges to the point symmetrical to $(a,c)$ with respect to $\ell^+$.
\end{itemize}
In each of these cases, the desired point $(a,c)$ is, respectively, $\mathrm{C}^-$, C, $\mathrm{C}^+$. 
The coordinates of $\mathrm{C}^{\pm}$ are given respectively by
\beq
\mathrm{C}^+: \frac{1}{(\epsilon^2 + 4)^{\frac{1}{2}}}(-\epsilon \alpha_c + 2 \gamma_c , 2 \alpha_c + \epsilon \gamma_c), \label{plus}
\eeq
\beq
\mathrm{C}^-: \frac{1}{(\epsilon^2 + 4)^{\frac{1}{2}}} (\epsilon \alpha_c - 2 \gamma_c , -2 \alpha_c - \epsilon \gamma_c). \label{minus}
\eeq
Because the point $(a,c)$ is on the region $a \ge 0$ (see (\ref{a_condition})), it is one of the points $\mathrm{C}$, $\mathrm{C}^+$ and $\mathrm{C}^-$ which is on the right-half plane of the $\alpha-\gamma$ plane. We can identify the coefficients $a$ and $c$ by investigating which point makes $y_1$ follow Gaussian distribution.

We summarize how we obtain $a$, $b$, and $c$.
\begin{enumerate}
\item After subtracting the means of the time series from them, transform $x_2(t)$ so that the equation (\ref{covariance}) is satisfied:
	\begin{equation}
	x_2(t) \to \frac{1}{\la x_2(t)^2 \ra^{\frac{1}{2}}} x_2(t).
	\end{equation}
\item Calculate $b_{est}$ by (\ref{b_est}).
\item Evolve $(\alpha, \gamma)$ following (\ref{alpha_evo2}) and (\ref{gamma_evo2}) while constraining the point on the region defined by (\ref{REGION}).
\item Calculate the coordinates of $\mathrm{C}^+$ and $\mathrm{C}^-$ by (\ref{plus}) and (\ref{minus}), and choose the best point that is on the right-half plane of $\alpha-\gamma$ plane and maximizes the Gaussianity of $y_1(t)$.
\end{enumerate}

\begin{figure}
	\begin{center}
		\includegraphics[width = 12cm]{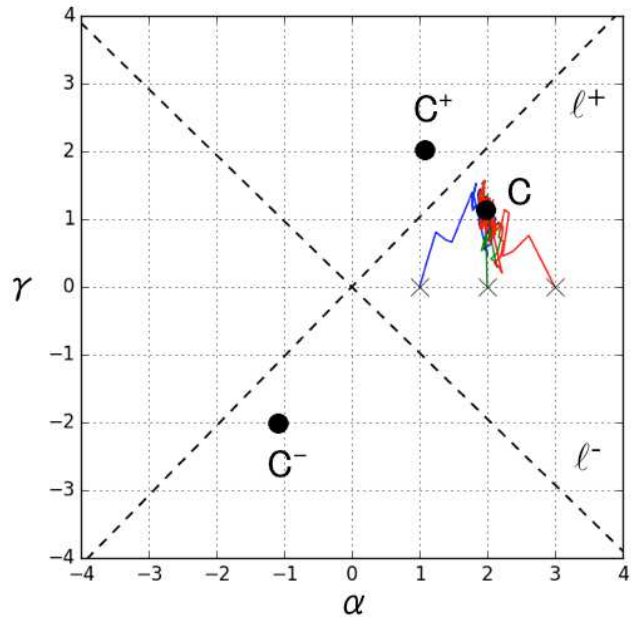}
		\caption{The evolutionary flows of (\ref{alpha_evo2}) and (\ref{gamma_evo2}) with $a=1$, $b=1$, $c=2$. The initial points are set to be (1,0), (2,0), (3,0). For each regeneration, we use only one sample to calculate the right hand side of (\ref{alpha_evo2}) and (\ref{gamma_evo2}), and do this procedure successively for each sample (This is called online algorithm.). For the $i$-th regeneration, the constant of proportionality is 0.01/(1 + $i$/1000), which begins to decrease at about the 1000th regeneration. The estimated value of $\epsilon$ is 0.05 and the lines, $\ell^{\pm}$, are drawn by using this value. As can be seen, the flows from all the initial locations converges to the same point. Because $\mathrm{C}^-$ is not on the right-half of the plane, the estimated point is $\mathrm{C}$ or $\mathrm{C}^+$. We use both values to construct $y_1$ according to (\ref{NLtransf}) and calculate each kurtosis of them, which is zero when the samples are normally distributed. We choose the point which makes the kurtosis smaller and find that the estimated values of $a$, $b$, and $c$ are $a_{est}=1.08$, $b_{est}=1.00$, and $c_{est}=2.01$.}
	\label{TEST}
	\end{center}
\end{figure}

In order to demonstrate that the algorithm works well, we make statistically independent samples of $x_1$ and $x_2$ (ignoring their temporal correlations) and test it. We assume that the probability density function of the non-Gaussian noise $k$ is the Laplace distribution function whose variance is unity and mean is zero:
\beq
q_2(k)=\frac{1}{\sqrt{2}} \mathrm{e}^{-\sqrt{2} |k|}.
\eeq
We generate 10,000 samples of $n$ and $k$, and calculate
\beq
x_1 = an+bk+cnk
\eeq
\beq
x_2 = k
\eeq
for each of the samples to make 10,000 samples of $x_1$ and $x_2$.
When we set $a=1$, $b=1$, $c=2$, the evolutionary flow on the $\alpha-\gamma$ plane is shown in Fig. \ref{TEST}.
The coordinates of $\mathrm{C}$, $\mathrm{C}^{\pm}$ are $\mathrm{C}(1.98, 1.13)$, $\mathrm{C}^+(1.08, 2.01)$ and $\mathrm{C}^-(-1.08,-2.01)$. The estimated point of $(a,c)$ is $\mathrm{C}$ or $\mathrm{C}^+$. 
We choose the point which makes the kurtosis of $y_1$ smaller, then find that the estimated values of $a$, $b$, and $c$ are $a_{est}=1.08$, $b_{est}=1.00$, and $c_{est}=2.01$, which are close to the real values.
 
Finally, we can remove the non-Gaussian noise $k(t)$ by using $y_1(t)$ instead of $x_1(t)$. By using (\ref{removeNG}), we can find that S/N is improved from
\beq
 \frac{S}{N}=\frac{h}{\sqrt{E[ n^2 ]+ \frac{b^2}{a^2} E[ k^2 ] + \frac{c^2}{a^2} (h^2 + E[ k^2 ]) E[ n^2 ]}}
\eeq
to
\beq
 \frac{S}{N}\approx\frac{h}{\sqrt{E[ n^2]}},
\eeq
as $y_1(t)$ reproduces $h(t)+n(t)$ well.

\section{Conclusion}
\label{conclusion}

In the present paper, we have provided a new way to handle non-Gaussian noises 
for the detection of GWs using the independent component analysis.  
While many other methods attempt to overcome the non-Gaussian nature of noises
by frontal attack, ICA makes use of non-Gaussianity as well as statistical independence,
to separate signals.

First, we have considered a simplified linear model to show that this method
may be useful to remove non-Gaussian noises that can be measured by
enviromental sensors at least partially. Next, we have checked the applicability of our method to
a more realistic case where there is time delay between the inputs and the outputs. We have shown that we can also remove non-Gaussian noises by
using Fourier components. 
These methods can be applied to the case which has
outputs of more channels.
Finally, we have tried to extend our method to nonlinearly 
correlated noises. 
As a first step, we have considered a specific nonlinear model and show that we can identify the coupling coefficients and remove non-Gaussian noises in that case.

It is worth while to pursue further
analysis by considering more realistic situations. 
Thus we shall prepare
for the completion of KAGRA detector toward gravitational wave astronomy.

\vskip 2cm
\noindent
{\large\bf Acknowledgements}

This work was partially supported by 
a research program of the Advanced Leading Graduate Course for Photon Science (ALPS) at the University of Tokyo (SM), the JSPS Grants-in-Aid for Scientific Research (KAKENHI) 15H02082(JY), 24103005 and 15K05070 (YI), and JSPS Fellows Grants No. 26.8636 (KE).

\end{document}